\documentclass[a4paper,british,english]{llncs}
\usepackage[T1]{fontenc}
\usepackage[latin9]{inputenc}
\usepackage{array}
\usepackage{amsmath}
\usepackage{graphicx}

\makeatletter

\providecommand{\tabularnewline}{\\}

\@ifundefined{showcaptionsetup}{}{%
 \PassOptionsToPackage{caption=false}{subfig}}
\usepackage{subfig}
\makeatother

\usepackage{babel}
\begin{document}
\selectlanguage{british}%

\title{A Tool to Estimate Roaming Behavior in Wireless Architectures}

\author{Rute Sofia}

\institute{COPELABS, University Lusófona, Lisboa, Portugal\thanks{\selectlanguage{english}%
COPELABS, Building U, First Floor, University Lusofona. Campo Grande
388, 1749-024 Lisboa, Portugal.\selectlanguage{british}%
}\linebreak{}
\email{rute.sofia@ulusofona.pt}}
\maketitle
\begin{abstract}
This paper describes a software-based tool that tracks mobile node
roaming and infers the time-to-handover as well as the preferential
handover target, based on behavior inference solely derived from regular
usage data captured in visited wireless networks. The paper presents
the tool architecture; computational background for mobility estimation;
operational guidelines concerning how the tool is being used to track
several aspects of roaming behavior in the context of wireless networks.
Target selection accuracy is validated having as baseline traces obtained
in realistic scenarios.
\end{abstract}


\section{Introduction}

The introduction of new, cooperative technologies and in particular
of low-cost wireless access, allowed the regular citizen to profit
from the Internet as a commodity. Such pervasive Internet access is
giving rise to networking architectures that seem to spread in a self-organizing
manner, \emph{User-centric Networks (UCNs)}~\cite{UPN,uloopbook}.
UCNs rely on an Internet end-user that exhibits frequent roaming patterns,
and that owns/carries one or more portable devices with a good multimedia
support. Hence, the majority of today's mobile devices, which have
been considered, up until recently, plain consumer devices, are now
also networking nodes. As a consequence, the movement that these devices
exhibit impact the underlying Internet connectivity models and the
overall network operation. Hence, being able to capture such movement
and also to estimate some features is highly relevant to optimize
different aspects of network operation e.g. resource management, or
routing.

Movement estimation has been a research field for long, with the aim
of improving network operation. For instance, in cellular networks,
several attempts have been provided to estimate movement as explained
in section 2. Fast handing over based on movement anticipation techniques
(e.g. tunneling) has been a topic extensively addressed, e.g., within
the context of the \emph{Internet Engineering Task Force (IETF)}. 

Estimating roaming behavior is therefore becoming more relevant, and
today, due to an extensive effort derived from several initiatives
as well as from extensive and wide traces collections \cite{crawdad},
it is globally accepted that there is a relation between social behavior
and the user's roaming behavior \cite{gonzalez2008understanding}.
It is the social behavior that assists in defining user movement patterns,
both from an individual perspective, and from a group perspective
\cite{rhee2011levy}. Being capable of estimating such behavior is
therefore relevant to optimize network operation, be it from a mobility
management perspective (e.g., handover optimization), from a resource
management perspective (e.g. performing a more intelligent load-balancing),
or from a routing perspective (e.g. making routing more stable by
selecting a priori paths that have a chance to be more stable in the
light of node movement).

This paper is dedicated to the topic of mobility prediction in wireless
networks. We provide a debate on notions related to social interaction
analysis and mobile networks as well as a debate on guidelines to
better address mobility prediction. Our work proposes a non-intrusive
wireless sensing tool, the MTracker\footnote{Software is available as a beta version available directly via Google
Apps, https://play.google.com/store/apps/details?id=eu.uloop.mobilitytracker,
via http://copelabs.ulusofona.pt/\textasciitilde{}uloop/ or at\\
 http://copelabs.ulusofona.pt/scicommons/index.php/publications/show/489.}, which provides a way to track properties of a user's visit to preferred
networks, and to estimate a potential move towards a more preferred
network, based on the learnt history of the user's roaming behavior. 

The paper is organized as follows. Section~\ref{sec:RelatedWork}
is dedicated to related work, while section \ref{sec:Ours} addresses
our proposed mechanism to estimate social mobility, and validate the
mechanism against real traces in section \ref{sec:validation}. The
paper concludes in section~\ref{sec:conclusions}, where guidelines
for future work are also provided.

\section{Related Work}

\label{sec:RelatedWork}

Within the context of cellular works, there are several studies dedicated
to movement prediction. First attempts related with prediction based
on \emph{Signal-to-Noise (SNR)} ratio levels~\cite{denkomobility},
being the main purpose to anticipate a potential handover and not
to predict such handover in terms, e.g., of preferential target. Improvements
to this line of research have been considered, for instance, via probabilistic
selection based on user \emph{Global Positioning System (GPS)} coordinates.
Such related work fell short in terms of adequately estimating movement,
partially due to the fact that at the time there was not a solid understanding
on users' roaming behavior.

The current availability of large-scale data sets, such as mobile-phone
records and GPS data, allows researchers from multiple scientific
fields to gain access to detailed patterns of human roaming behavior,
greatly enhancing our understanding of human mobility. The extensive
traces that are available today lead to a better understanding of
social movement, having given rise to a few mobility models with roots
on social network theory~\cite{musolesi4,MySurvey}. 

In terms of human movement, Barabási et al. have been active in giving
insight into human movement patterns~\cite{understanding}. As follow-up
of their research, Song et al. research showed that human movement
behavior is not compatible with Brownian approaches and showing also
some level of predictability in such movement~\cite{SONG}. By measuring
the entropy of each individual's trajectory, they have found that
there is a 93\% predictable behavior across the studied universe.

Noulas et al. have analyzed roaming behavior features exploiting information
on transitions between types of places, mobility flows between venues,
and spatio-temporal characteristics of user check-in patterns, showing
that supervised models based on the combination of multiple features
assist in reaching high prediction accuracy \cite{noulas:mining}.
Their analysis is focused on mobility prediction targetting location-based
services. Our work has in common with the latter the intention to
consider social behavior aspects that can be extracted from visits
to networks, to improve mobility prediction. We do not, however, consider
location-based services as the target to address. Instead, our perspective
is derived from data captured passively by the user device only, in
a non-intrusive way.

\section{A Tool to Estimate Movement in Wireless Networks}

\label{sec:Ours}

The MTracker solution is a proof-of-concept software-based mechanism
that intends to optimize wireless networking in the following aspects:
i) handover optimization by improving resulting end-to-end delay (and
node reachability time), as well as by reducing signaling overhead
associated to handovers; ii) optimize resource management by estimating
potential attachment points, and assisting the network in self-organizing,
thus providing stations with the optimal wireless base stations.

\subsection{Functional Aspects}

There are three main tasks that the MTracker performs \cite{IPRMTracker}.
The first is\textbf{ Data Capture}, namely, non-intrusive data capture
based on visited and surrounding wireless networks. The second is
\textbf{Target Handover Selection}, i.e., preferential target selection
based on a seamless ranking of all wireless networks on a device's
list. The third one, \textbf{Time-to-Handover}, concerns estimating
the time for the next handover to occur, based on the learnt roaming
behavior of the device.

Fig. \ref{fig:MTracker-Flow-chart-in} provides a flow-chart for the
MTracker operation. Its main three tasks are explained next.

\subsubsection{Data Capture}

\begin{figure}
\includegraphics[scale=0.42]{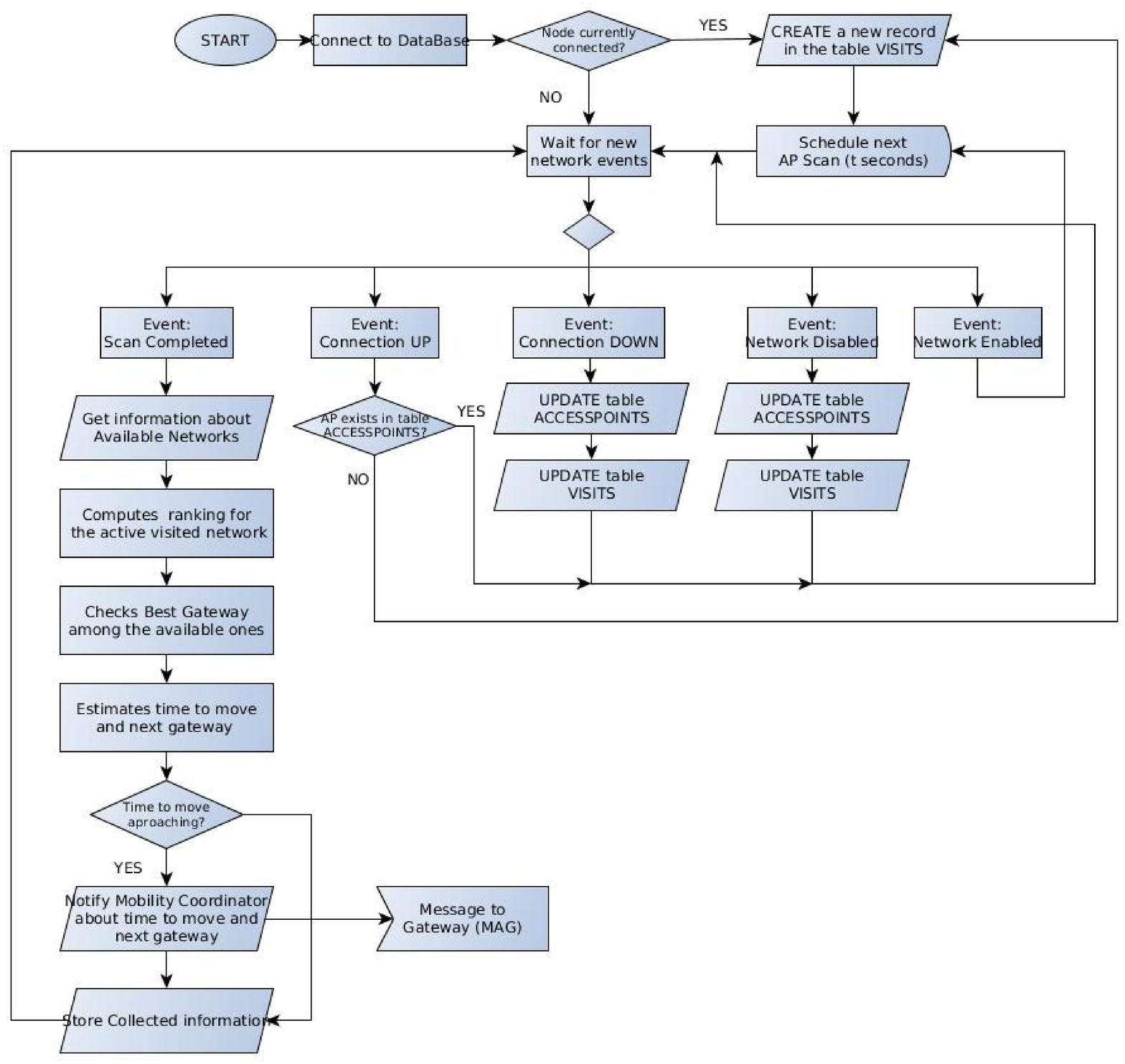}

\protect\caption{\selectlanguage{english}%
MTracker Flow-chart .\label{fig:MTracker-Flow-chart-in}\selectlanguage{british}%
}
\end{figure}

Once the MTracker is activated, it relies on the usual 802.11 MAC
Layer scanning mechanism to periodically obtain data concerning the
list of networks in range, as illustrated in Fig. \ref{fig:MTracker-Flow-chart-in}. 

The MTracker list of visited networks is kept locally on the device.
The tool devised is capable of periodically provides output to a local
system (e.g. local database); or to external entities that reside
on the network, e.g. a mobility management solution that has the responsibility
to decide based on other external conditions whether or not a move
should be anticipated. 

The MTracker captures parameters via the Wi-Fi interface that are
either overheard or that can be computed based on overhead data \nobreakdash-
it does not perform intrusive probing. The initial set of parameters
considered in the MTracker are described in Table \ref{tab:Initial-set-of}\footnote{The MTracker as proof-of-concept has already given rise to the tool
WiRank (https://play.google.com/store/apps/details?id=eu.uloop.wirank),
intended to improve Android connectivity and to integrate some aspects
of prediction with other features, such as context-awareness based
on location.}.

\begin{table}
\protect\caption{\selectlanguage{english}%
Some parameters collected by the MTracker.\label{tab:Initial-set-of}\selectlanguage{british}%
}

{\scriptsize{}}%
\begin{tabular}[t]{|>{\raggedright}p{1.5cm}|>{\raggedright}p{1cm}|>{\raggedright}p{3cm}|}
\hline 
\textbf{\scriptsize{}Parameter} & \textbf{\scriptsize{}Name} & \textbf{\scriptsize{}Definition}\tabularnewline
\hline 
{\scriptsize{}$v_{ij}\in[0,\propto]$} & \selectlanguage{english}%
{\scriptsize{}visit}\selectlanguage{british}%
 & {\scriptsize{}A visit from node i to node j implies that node i is
authorized (by j) to use its networking resources.}\tabularnewline
\hline 
{\scriptsize{}${\textstyle v=\sum_{j=0}^{n}v_{ij}\ j\in[0,n]}$} & \selectlanguage{english}%
{\scriptsize{}Total visits}\selectlanguage{british}%
 & {\scriptsize{}Number of visits that node i does to node j.}\tabularnewline
\hline 
{\scriptsize{}$d_{ij}$} & {\scriptsize{}Visit duration} & {\scriptsize{}Time interval (seconds) since node $i$ is authorized
by node j to be attached, until node i deattaches.}\tabularnewline
\hline 
{\scriptsize{}$d_{avg}=\gamma*d_{avg}+(1-\gamma)*d_{ij}$} & {\scriptsize{}average duration of a visit} & {\scriptsize{}Time interval (seconds) that node i is in average attached
to node j, based on nan exponential moving average formula.}\tabularnewline
\hline 
{\scriptsize{}$a_{ij}$} & {\scriptsize{}visited network attractiveness} & {\scriptsize{}A parameter that a user sets by hand (e.g. gives more
preference to using network1 than network2) or it can be passively
collected via, e.g., distributed trust schemes that are present in
the network (e.g. provided by the operator).}\tabularnewline
\hline 
\selectlanguage{english}%
{\scriptsize{}$rej_{ij}$}\selectlanguage{british}%
 & {\scriptsize{}Rejected visits} & {\scriptsize{}Number of times a node i is not authorized by node j
to access its resources.}\tabularnewline
\hline 
\selectlanguage{english}%
{\scriptsize{}$te_{ij}$}\selectlanguage{british}%
 & {\scriptsize{}Visit gap} & {\scriptsize{}Time gap (in seconds) since the last visit from node
i to a specific visited network j.}\tabularnewline
\hline 
\end{tabular}{\scriptsize \par}
\end{table}

The mentioned parameters are here provided as a potential example
of the type of parameters that can be used to characterize a user\textquoteright s
roaming behavior in terms of preferred networks without recurring
to probing or to explicit location tracking. These parameters are
used to compute the visited networks' rank, as explained next.

\subsubsection{Target Handover Selection, Ranking Visited Networks}

The MTracker tool has been designed to integrate any utility function
to rank visited networks. In this paper we consider a potential equation,
Eq. \ref{eq:1-1-1}, where $r_{ij}$ corresponds to the ranking (cost)
that node $i$ computes towards the network controlled by node $j$.
The rationale for such equation is that the longer and the more often
a node visits a specific network, the higher the preference of that
network to the node, provided that such visits are recent. Hence,
the function described in Eq. \ref{eq:1-1-1} has enough sensitivity
to distinguish between targets that seem to be preferential (for instance,
high $a_{ij}$ and long $d_{avg}$) but that have actually been heavily
visited a long time ago (long $te_{ij}$). The function also takes
into consideration the number of rejected connections $rej_{ij}$
against the total number of visits $v_{ij}$.

The rank provided by $r_{ij}$ is computed from the perspective of
node $i$ towards a potential visited network identified by node $j$
(e.g. an AP) based on parameters passively collected over time, by
relying on the exponential moving average function of Eq. \ref{eq:rEMA},
where $r_{ij_{t-1}}$corresponds to the last computed value for $r_{ij}$and
$r'_{ij}$stands for the instant computation of $r_{ij}.$ By tuning
$\alpha$ one shall be providing more weight to more recent or to
older instances of $r_{ij}$.

\begin{equation}
r_{ij}=a_{ij}^{2}*(\frac{\sqrt{d_{avg}}}{te_{ij}+1})^{\frac{v}{rej_{ij}}}\:a_{ij}\:\in[0,1]\label{eq:1-1-1}
\end{equation}

\begin{equation}
r{}_{ij}=\alpha*r_{ij_{t-1}}+(1-\alpha)*r'_{ij},\,\alpha\:\in[0,1]\label{eq:rEMA}
\end{equation}

\subsubsection{Time-to-Handover Estimate}

Estimating a potential move is a task processed by a node in background
and has as motivation to provide an estimate of time, as well as a
target identifier for the next handover to be performed by the node.
We highlight that the MTracker only notifies an entity (a user, some
entity on the network, or even some other process in the local device)
that a potential move may occur, so that a decision may assist the
device in reaching some form of reliability in terms of active communication
flow. For instance, it is still up to a mobility management solution
to perform such a move, or not, based on the information provided
by the MTracker.

To compute the estimate for a potential move, the MTracker periodically
checks its list of ranked visited networks. Based on the computed
average visit time of the active network as well as on the error time
gap derived from prior learning about roaming habits of the node,
the MTracker verifies which network(s) attain the best ranking in
comparison to the active network.

The time to handover, \emph{TTH}, estimated during an active connection
of node $i$ to node $j$ is based on Eq. \ref{eq:1-1-1-1}. The equation
takes into consideration the average visit duration to the network
controlled by node $j$, $d_{avg}$ as well as the time gap $\triangle t$.

\begin{equation}
TTH_{t}=d_{avg}\pm\triangle t,\:where\:\triangle t=d_{t-1}-TTH_{t-1}\label{eq:1-1-1-1}
\end{equation}

\subsection{Operational Example}

\begin{figure}
\center\includegraphics[scale=0.4]{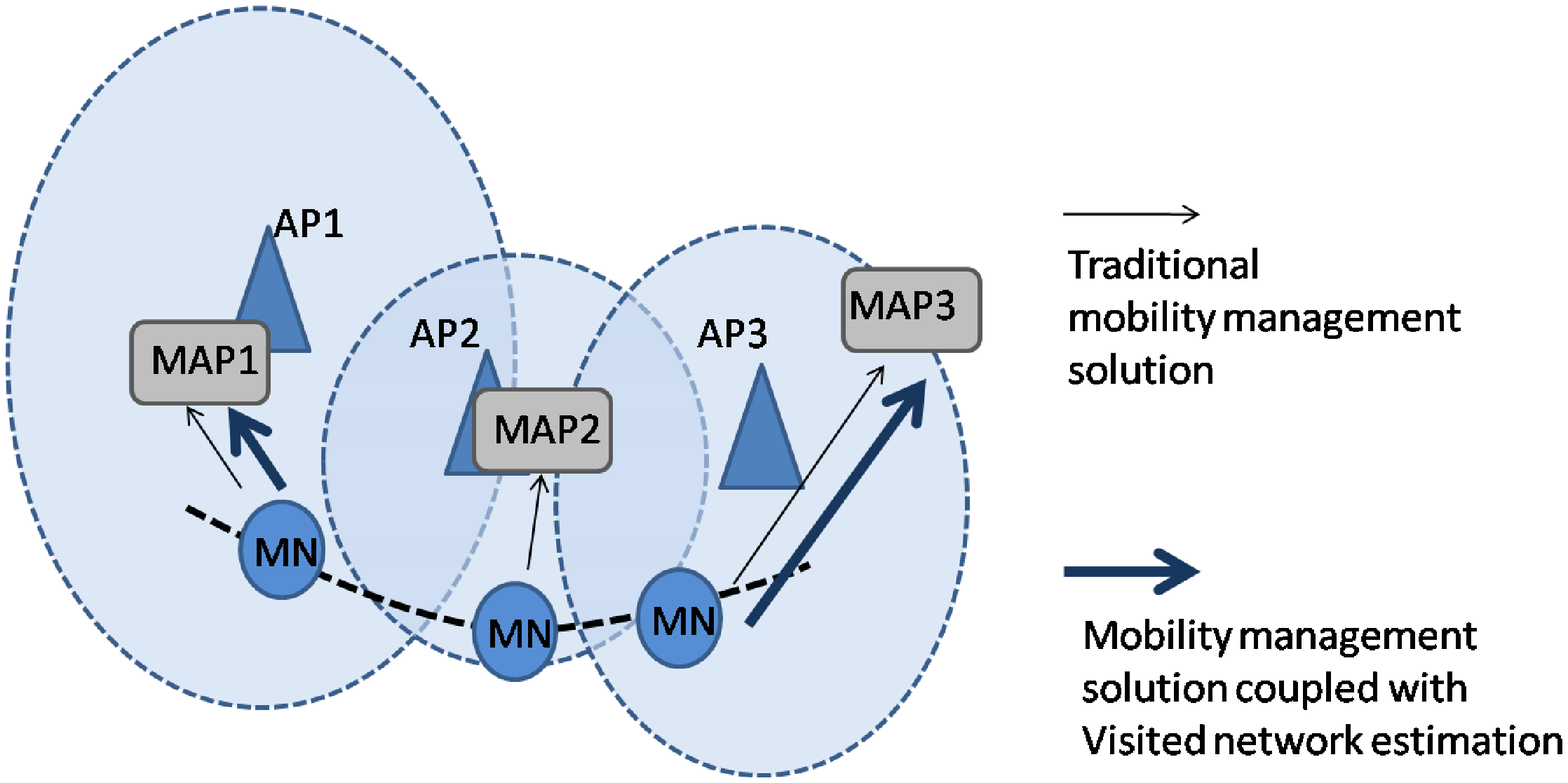}

\protect\caption{\selectlanguage{english}%
\label{fig:Example-of-Impact-1}Example of the potential impact of
mobility estimation f.\selectlanguage{british}%
}
\end{figure}

Fig. \ref{fig:Example-of-Impact-1} illustrates a wireless scenario
where three wireless visited networks are respectively served by AP1,
AP2, and AP3. The application MTracker resides on the \emph{Mobile
Node (MN)}, which periodically visits the three networks. Moreover,
each visited network is served by a specific \emph{Mobility Anchor
Point (MAP)} agent which can be co-located to the AP, or placed somewhere
else on the network, as occurs today. 

MN exhibits a regular trajectory e.g. during a day, where it crosses
the three different visited networks. Following the regular IEEE 802.11
operation MN is set to perform \emph{passive scan}ning, i.e., while
roaming it passively receives \emph{Beacon} frames sent by the surrounding
APs. It can therefore get a list not only of APs that it regularly
attaches to, but also of neighboring APs that it did not visit. We
highlight that there is no relation whatsoever with GPS location or
tracking of the nodes; the MTracker simply captures data that is already
provided by today's devices, and has the capability to infer roaming
behavior in terms of characteristics for the next handover.

In this operational example, intended to illustrate the benefits of
prediction in mobility management, MN also integrates a mobility management
solution, e.g. \emph{Mobile IPv6} (MIPv6) or \emph{Proxy Mobile IPv6}
(PMIPv6). On its list of visited locations, it keeps track of multiple
parameters such as the ones described in Table 1. 

For instance, MN has recorded an average visit duration ($d_{avg}$)
of 15 minutes to AP1. On the current visit, 6 minutes have elapsed.
Periodically, MN analysis its list of visited networks and checks
whether or not the average duration visit is being reached. From a
computational perspective, this means that MN integrates a time-window
based mechanism to reach and eventually send a notification to an
entity in the current visited network (e.g., AP, MAP, etc), e.g. every
minute. 

In our example, after 6 minutes, MN realizes that there is still a
gap of 9 minutes and therefore does not send any information. When
MN1 realizes that the current visit has reached 14 minutes, it sends
a notification about the best possible visited network which, in our
example, is the visited network served by AP3. In that notification,
it therefore sends information to MAP1 about the best next MAP \textendash{}
MAP3, and also about how much time in average is left for a move.
MN does not perform, however, any decision concerning moving (handover). 

This outcome can then be fed to a mobility management process, which
can then decide whether or not to activate a handover, as we have
addressed in prior work \cite{uloopbook}. However, this outcome can
also be fed into other control-based processes as a way to estimate
aspects of roaming behavior.

\section{Target Selection Accuracy Validation}

\label{sec:validation}

In this section we provide input concerning the performance of mobility
prediction based on an analysis performed by considering realistic
traces. The validation contemplates the \textbf{accuracy} of the tool
in terms of adequately ranking preferencial networks. The traces selected
are from University of South Carolina and available via the CRAWDAD
repository \cite{crawdad}. The full set of USC traces \footnote{http://uk.crawdad.org/usc/mobilib/.}
comprise 150 nodes which have been tracked over several months across
different visited networks. Extracted traces provide information such
as identifier of each visited network; duration of each visit; timestamp
for the visit start. As there are no traces that consider rejection
rates as well as attractiveness level, we have considered attractiveness
to be similar to the number of visits $v$ and did not assess the
impact of rejected visits in our experiments.

\begin{figure}
\subfloat[\selectlanguage{english}%
MN45 Trajectory.\selectlanguage{british}%
]{\protect\includegraphics[scale=0.2]{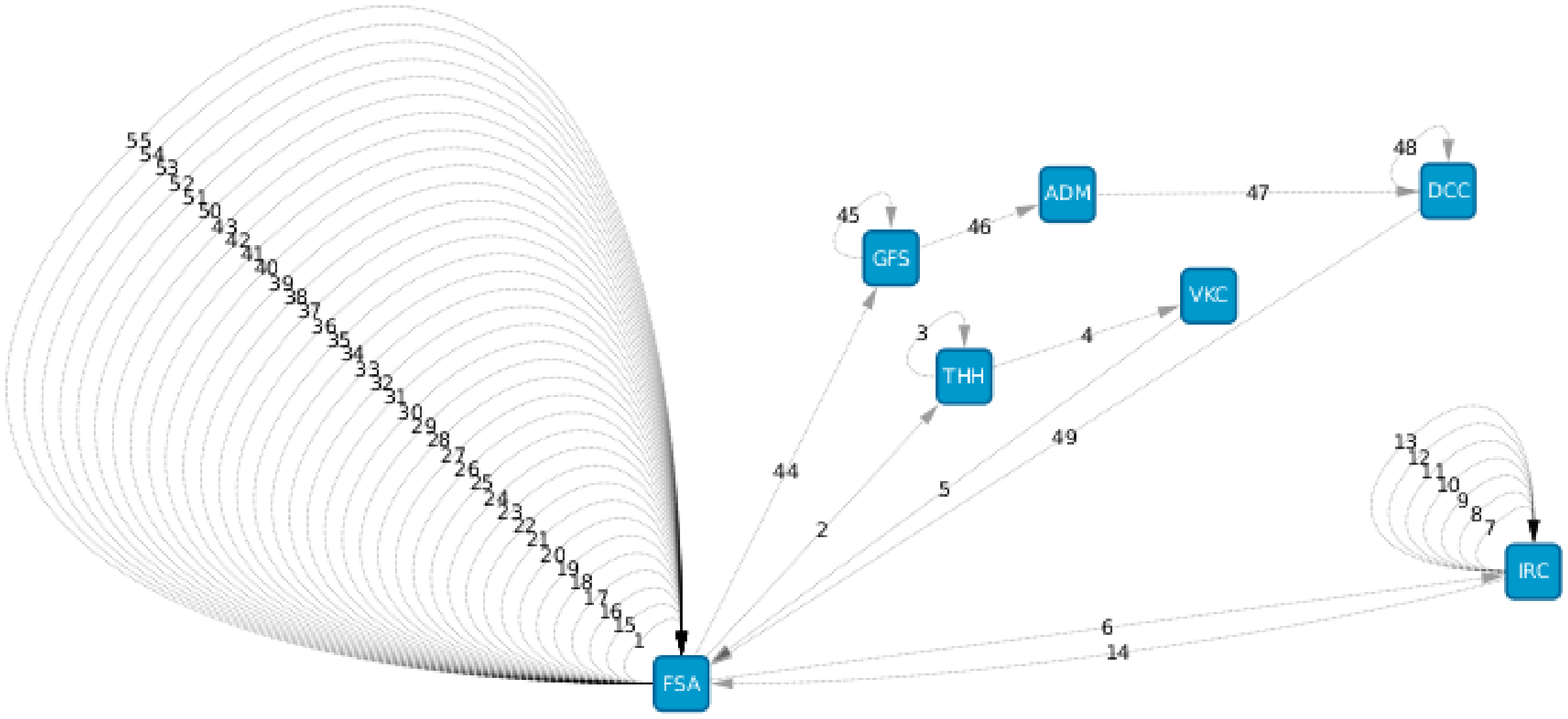}

}\subfloat[\selectlanguage{english}%
MN36 Trajectory.\selectlanguage{british}%
]{\protect\includegraphics[scale=0.2]{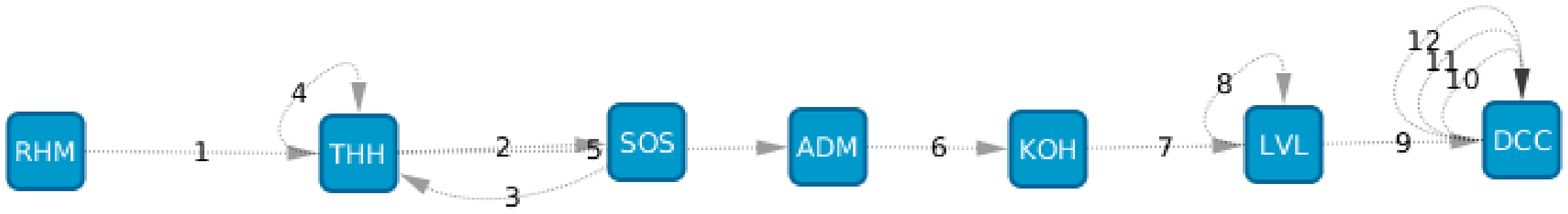}

}\\

\subfloat[\selectlanguage{english}%
MN35 Trajectory.\selectlanguage{british}%
]{\protect\includegraphics[scale=0.2]{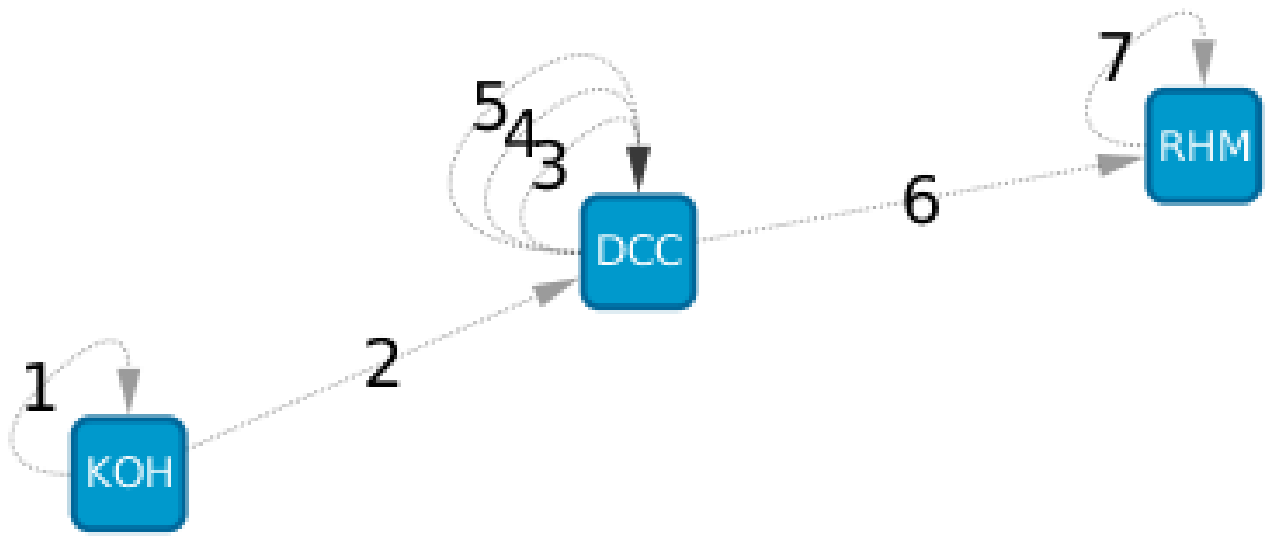}

}\subfloat[\selectlanguage{english}%
MN28 Trajectory.\selectlanguage{british}%
]{\protect\includegraphics[scale=0.2]{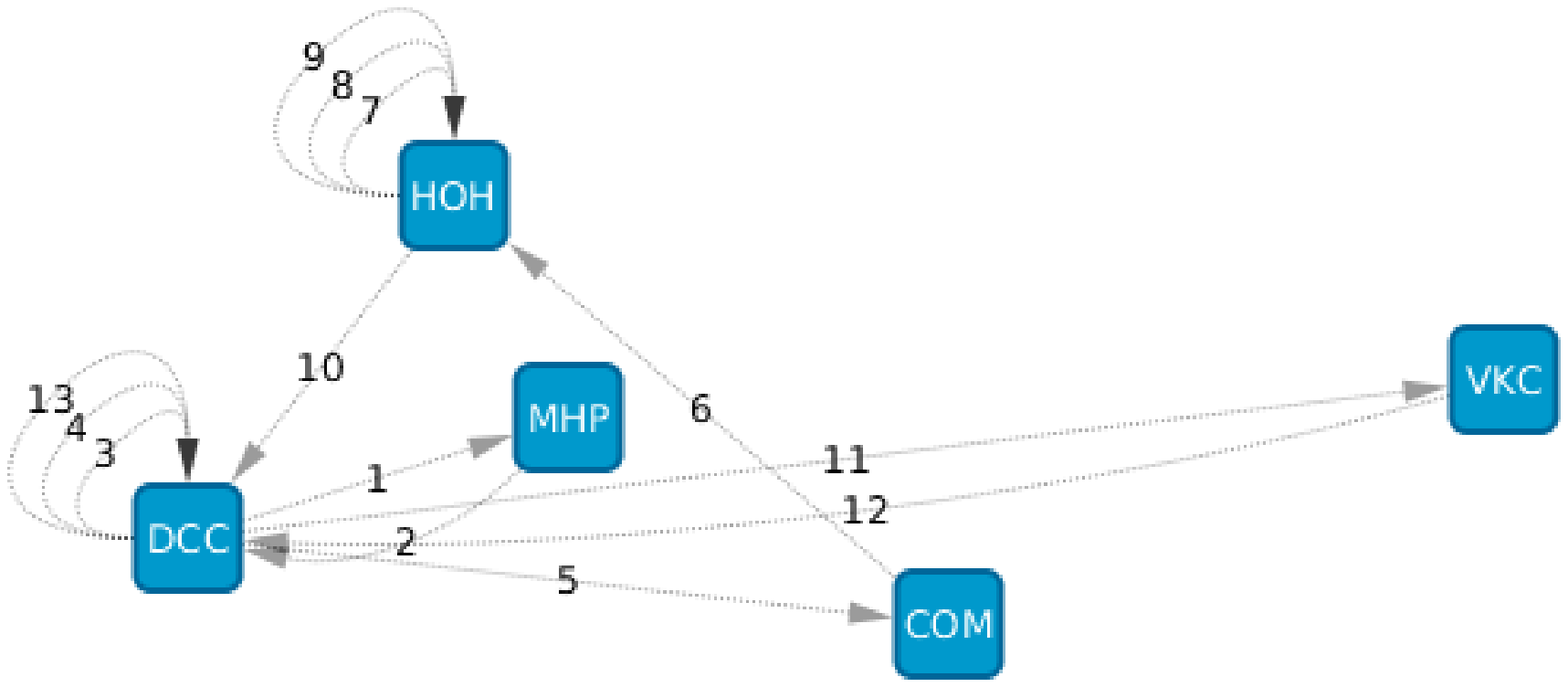}

}\\

\subfloat[\selectlanguage{english}%
MN90 Trajectory.\selectlanguage{british}%
]{\protect\includegraphics[scale=0.2]{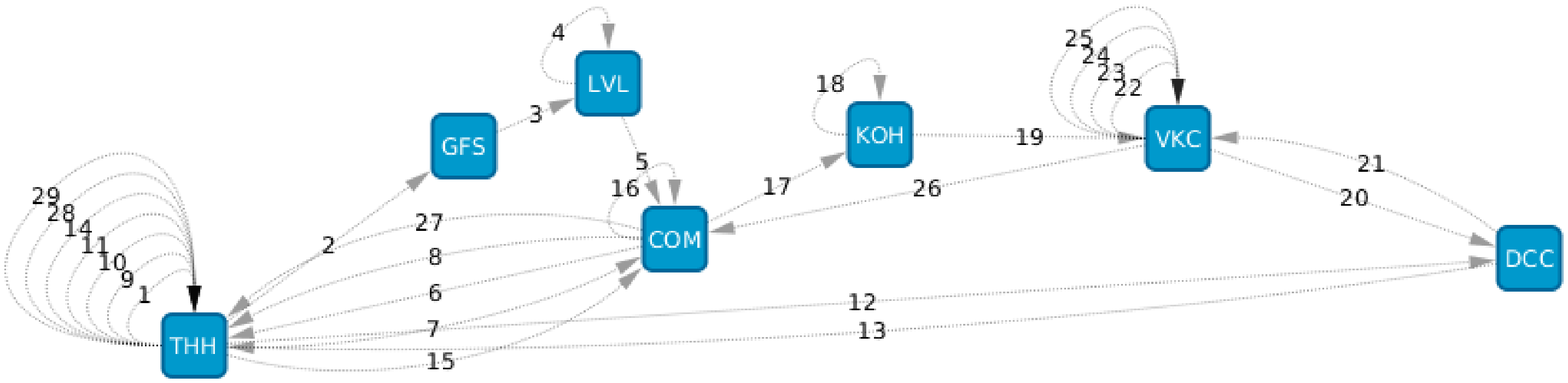}

}\subfloat[\selectlanguage{english}%
MN14 Trajectory.\selectlanguage{british}%
]{\protect\includegraphics[scale=0.2]{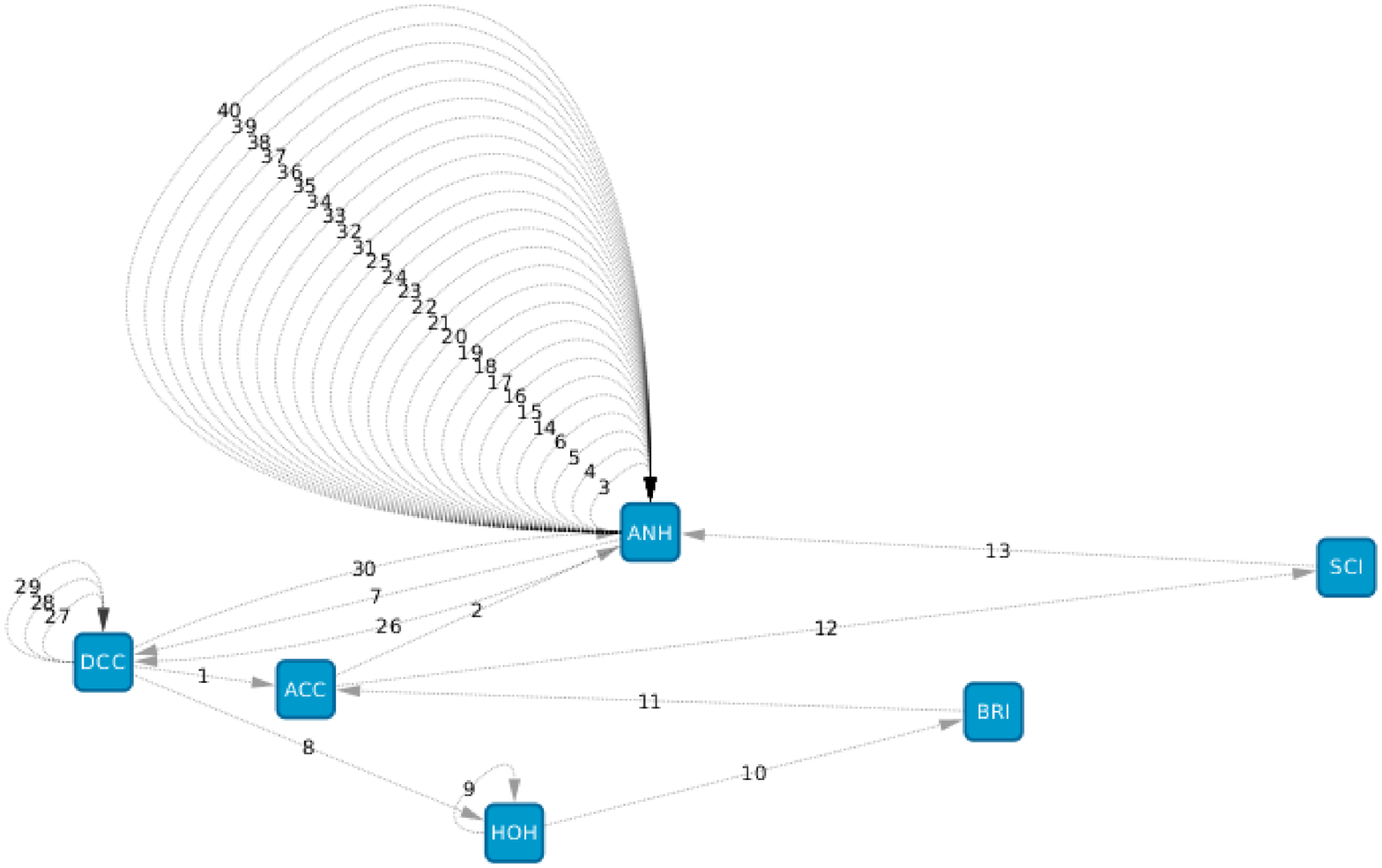}

}\\

\subfloat[\selectlanguage{english}%
MN21 Trajectory.\selectlanguage{british}%
]{\protect\includegraphics[scale=0.2]{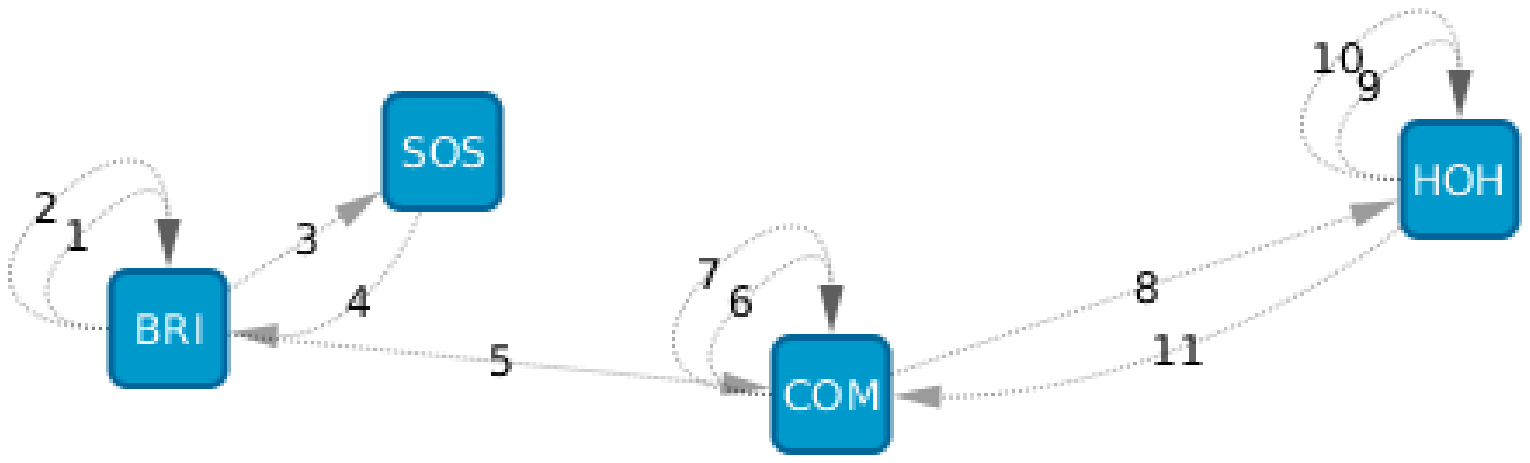}

}\subfloat[\selectlanguage{english}%
MN29 Trajectory.\selectlanguage{british}%
]{\protect\includegraphics[scale=0.2]{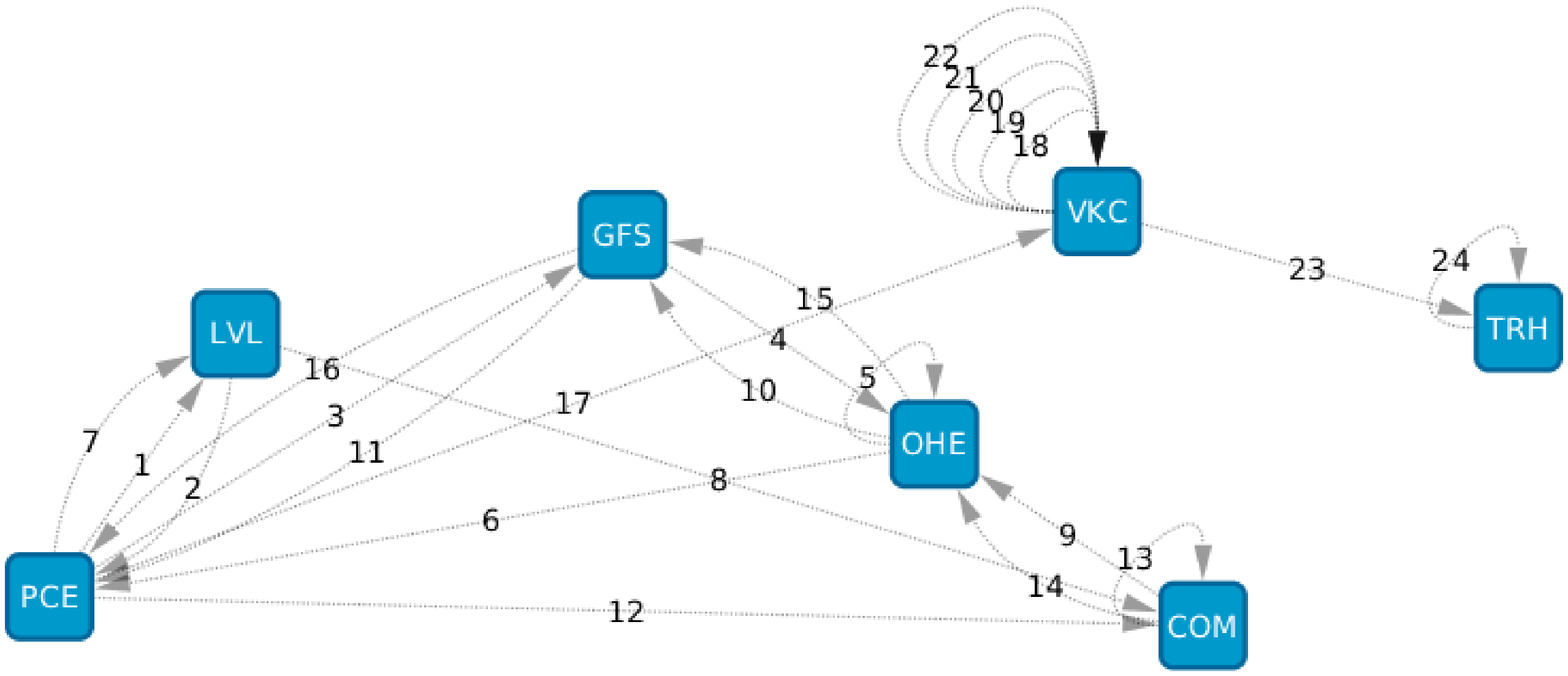}

}\\

\subfloat[\selectlanguage{english}%
MN34 Trajectory.\selectlanguage{british}%
]{\protect\includegraphics[scale=0.2]{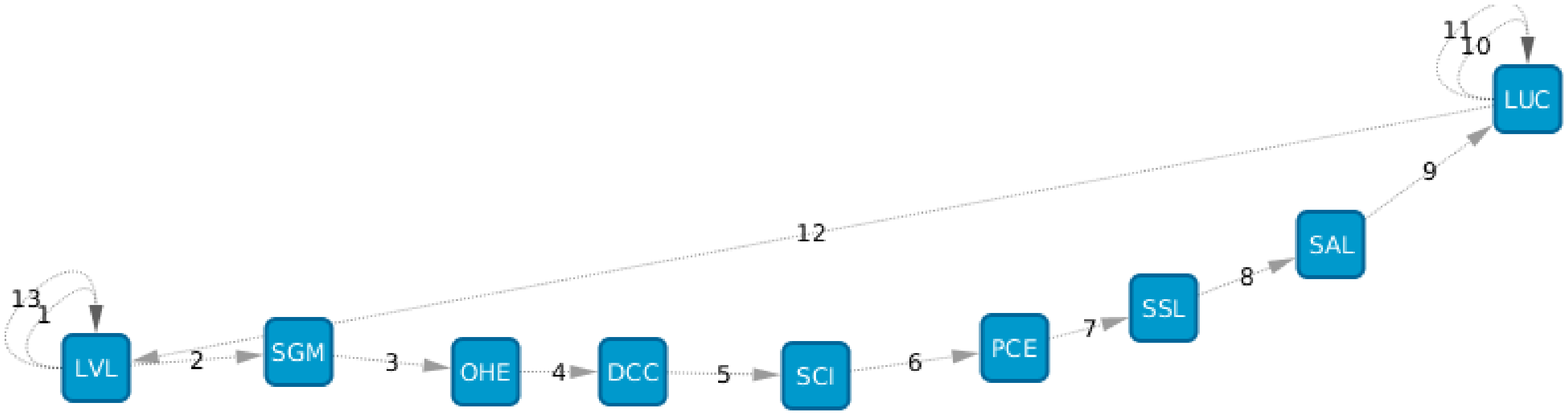}

}

\protect\caption{\selectlanguage{english}%
\label{fig: Trajectories}Trajectories of the nine selected nodes,
MN45, MN36, MN35, MN28, MN90, MN14, MN21, MN29, MN34. Edges numbering
refers to waypoints.\selectlanguage{british}%
}
\end{figure}

Out of the available traces, we have categorized nodes in terms of
trajectory, namely: trajectory duration (short; long); number of visited
APs (small number, large number); number of waypoints (high or low
number of waypoints in the trajectory). Then, for the different categories,
we have randomly selected again nine representative node trajectories.
These nodes, numbered respectively as MN45, MN36, MN28, MN90, MN14,
MN21, MN29, MN34, have their trajectories represented in Fig. \ref{fig: Trajectories}. 

The ranking accuracy has been validated by computing the ranking of
each visited AP over time and then selecting the AP with the highest
rank, among all possible APs. 

Table \ref{tab:Results,-ranking-accuracy} provides results for the
set of selected nodes. The first column represents the selected nodes.
The second column contains the total of visited networks for that
node during a single trajectory, while the third column provides the
total number of visits across different networks. The fourth column
provides details concerning the roaming paths extracted of the traces
for each node, while the fifth and six columns provide details concerning
the average visit duration (seconds and minutes) and total duration
of the traces (seconds and days), respectively. The relative error
percentage is provided in column seven. 

\begin{table*}
\protect\caption{\selectlanguage{english}%
\label{tab:Results,-ranking-accuracy}Results, ranking accuracy based
on movement of 9 nodes collected in the USC CRAWDAD trace set.\selectlanguage{british}%
}

\center%
\begin{tabular}{|>{\raggedright}p{1cm}|>{\centering}p{1cm}|>{\centering}p{1cm}|>{\centering}p{3cm}|>{\centering}p{2cm}|>{\centering}p{2cm}|>{\centering}p{1cm}|}
\hline 
\textbf{\tiny{}MN} & \textbf{\tiny{}Number of visited APs} & \textbf{\tiny{}Total waypoints} & \textbf{\tiny{}Roaming path main features} & \textbf{\tiny{}Average visit duration (seconds/minutes)} & \textbf{\tiny{}Total roaming duration (s/days)} & \textbf{\tiny{}Ranking accuracy error margin (\%)}\tabularnewline
\hline 
\hline 
{\scriptsize{}MN45} & {\scriptsize{}7} & {\scriptsize{}55} & {\scriptsize{}1 AP heavily visited; remainder APs visited in average
3 times} & {\scriptsize{}2793 / 46} & {\scriptsize{}621978/7 } & {\scriptsize{}20}\tabularnewline
\hline 
{\scriptsize{}MN36} & {\scriptsize{}6} & {\scriptsize{}12} & {\scriptsize{}Most APs visited once; 2 visited more than twice} & {\scriptsize{}2112/35} & {\scriptsize{}1656107/19} & {\scriptsize{}25}\tabularnewline
\hline 
{\scriptsize{}MN35} & {\scriptsize{}3} & {\scriptsize{}8} & {\scriptsize{}2 APs visited twice; 1 visited 4 times} & {\scriptsize{}162757/2712} & {\scriptsize{}1303008/15} & {\scriptsize{}38}\tabularnewline
\hline 
{\scriptsize{}MN28} & {\scriptsize{}5} & {\scriptsize{}14} & {\scriptsize{}3 APs visited once only} & {\scriptsize{}8336/139} & {\scriptsize{}3334673/38} & {\scriptsize{}43}\tabularnewline
\hline 
{\scriptsize{}MN90} & {\scriptsize{}7} & {\scriptsize{}29} & {\scriptsize{}All APs visited at least twice; Most APs visited frequently
for a specific timeslot, but not revisited} & {\scriptsize{}1908/31} & {\scriptsize{}1295067/15} & {\scriptsize{}54}\tabularnewline
\hline 
{\scriptsize{}MN14} & {\scriptsize{}6} & {\scriptsize{}41} & {\scriptsize{}1 AP accounts for circa 50\% of visits; all APs revisited} & {\scriptsize{}9250 / 154} & {\scriptsize{}1346709 / 16} & {\scriptsize{}29}\tabularnewline
\hline 
{\scriptsize{}MN21} & {\scriptsize{}4} & {\scriptsize{}12} & {\scriptsize{}1 AP visited only once; all APs revisited at least 3
times sequentially} & {\scriptsize{}5803 / 98} & {\scriptsize{}366220 / 4} & {\scriptsize{}33}\tabularnewline
\hline 
{\scriptsize{}MN29} & {\scriptsize{}7} & {\scriptsize{}24} & {\scriptsize{}Most APs visited twice only; 1 AP visited often} & {\scriptsize{}3819 / 64} & {\scriptsize{}1225566 / 14} & {\scriptsize{}28\%}\tabularnewline
\hline 
{\scriptsize{}MN34} & {\scriptsize{}9} & {\scriptsize{}14} & {\scriptsize{}7 APs visited only once} & {\scriptsize{}1993 / 33} & {\scriptsize{}949480 / 11} & {\scriptsize{}7.14}\tabularnewline
\hline 
\end{tabular}
\end{table*}

MN45 stands for an example of a node that exhibits a long trajectory
with a high number of waypoints, across a small number of visited
networks (7). MN45 exhibits frequent visits to two different visited
networks, represented as FSA and IRC. These are regular visits over
time. In this case, the ranking capability provides an error estimate
of around 20\%, which is quite relevant given the fact that the other
visited networks are in average visited only 3 times, are not necessarily
visited sequentially.

MN36 stands for an example of a node that shares a similar trajectory
in terms of time and visited networks, with the difference that its
trajectory holds less waypoints in comparison to MN45 \nobreakdash-
12 instead of 55. For this case, the accuracy is similar, as even
though the visited networks have in average been less visited. The
reason for this is that visits have been in average longer, which
compensates for the lack of visit frequency.

MN28 and MN90 are worst-cases in terms of predicting next handover
targets, as the MTracker has reached respectively an error rate of
43\% and 54\%. The reason relates to the fact that only two networks
are visited more than once, with short visits. 

MN34 stands for the best-case in terms of target accuracy (7\% of
error margin), being the reason the fact that only the first and the
last networks of the trajectory have been frequently and sequentially
visited. These networks also exhibit in average longer visits than
the others. 

The results obtained show that the MTracker can predict with reasonable
accuracy future handover targets, as the error margin stands between
20\% and 30\%. 

\begin{figure}
\center\subfloat[\selectlanguage{english}%
\label{fig:rankerror}Error margin across all cases.\selectlanguage{british}%
]{\protect\includegraphics[scale=0.6]{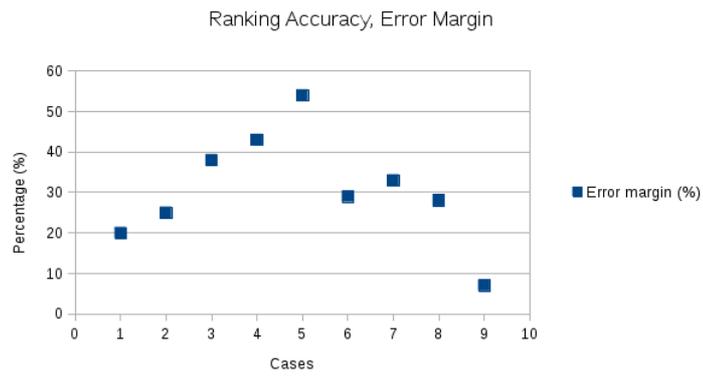}

}

\subfloat[\selectlanguage{english}%
\label{fig:CDF}Probability Density and Cumulative Density.\selectlanguage{british}%
]{\protect\includegraphics[scale=0.6]{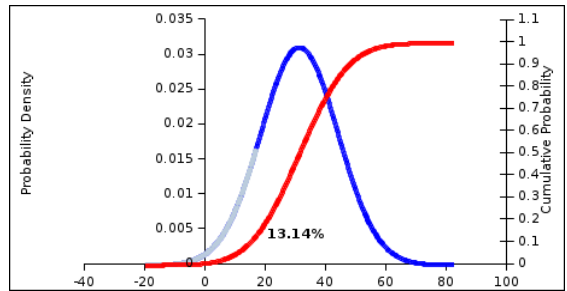}

}

\protect\caption{\selectlanguage{english}%
Ranking error margin.\selectlanguage{british}%
}
\end{figure}

The error margins obtained accross all cases are provided in Fig.\ref{fig:rankerror}
and Fig. \ref{fig:CDF} (Cumulative Density and Probabilistic Density
Functions). 

\section{Summary and Conclusions}

\label{sec:conclusions}

This paper contributes to the debate concerning the validity of mobility
estimation as an operational improvement to networks, and one that
can be easily implemented. It goes over the main concepts concerning
movement estimation for portable devices, in particular in UCNs, where
the end-user device often assumes the role of a network device. 

The work describes the tool MTracker explaining how it can be used
to assist in predicting roaming behavior in terms of handover target
and time to handover. 

We have validated the tool against traces obtained from devices realistic
environments, showing that without considering GPS, MTracker is capable
of predicting with reasonable accuracy future targets (20-30\% error
margin for most cases).

As current work, our mobility estimation framework is being validated
in the context of IP mobility management solutions. It is also being
extended to capture additional parameters that may assist other aspects
of the network operation, e.g., routing. Traces are starting to be
collected and will be provided to the global community. Last, but
not the least, mobility estimation aspects are also being used to
feed visualizing tools online, as a way to further analyze mobility
behavior of Internet users, in a seamless, anonymous and yet pervasive
way.

\section*{Acknowledgments}

This work has been developed within the EU IST FP7 project \emph{ULOOP
- User-centric Wireless Local Loop} (grant number 247158). The author
thanks Jonnahtan Saltarin for the implementation of the second version
of the proof-of-concept software MTracker.

\bibliographystyle{elsarticle-num}
\bibliography{rcl_rs}
\selectlanguage{english}%

\end{document}